# Equivalence of macro- (Van der waals) and micro- (quantum) phase-transitions: consequences for BEC, antihydrogen and generic chiral symmetry breaking

G. Van Hooydonk, Ghent University, Faculty of Sciences, Krijgslaan 281, B-9000 Ghent (Belgium)

**Abstract**. The internal microscopic phase transition in atom H described with a Hund-type Mexican hat potential is analytically congruent with the macroscopic phase transitions between states of aggregation, described with classical Van der waals-Maxwell binodals and spinodals. This means the UEOS (universal equation of state) exists and is valid for any phase transition in any system. A universal chiral symmetry breaking mechanism applies to BEC, where the *de Broglie* length is important. We show quantitatively that his standing wave equation leads to the equivalence of macro- and microscopic phase transitions, due to natural chiral symmetry breaking. The experimentally detected phase transition in the so-called simple Coulomb electron-proton bond between the 2 enantiomers H and antiH is the generic prototype.

## 1. Introduction

In line with recent results on phase transitions in small (microscopic) neutral systems due to charge-inversion [1a], we can now report on a striking similarity these microscopic (quantum) phase transitions and those known for centuries in natural stable macroscopic systems. As stated in [1a], the UEOS (universal equation of state) will have to apply for all details (phase transitions, critical points) of macro- [1b] and micro-systems, even BEC [1c], as discussed in [1a]. Already in the 19$^{th}$ century, Van der waals' EOS [2a] rationalized macroscopic phase transitions[1] between *states of aggregation*, but for a micro-system, *states of aggregation are meaningless*: his state qualifications like solid, liquid and gas do not apply for a single unit. Still, macro-behavior is *species-specific:* a unit of a *species* must already contain all information, needed for its macro-system to (re-)organize. Gibbs' phase rule links degrees of freedom f with c components and p phases by f =c –p +2, which makes the triple point for *the 3 classical macro-phases (solid-liquid-gas) of a species invariant*. But *internal phases* and quantum effects within a micro unit must also be assessed by the UEOS, which makes it so difficult to find it [1,3-4]. Atomic and molecular spectra show *internal phases occurring in single neutral units*. Other phenomena like superfluïdity, superconductivity (BCS) and BEC [5] point to more phases than classical *states of aggregation* and than states revealed by spectra. Theoretically, Van der waals-interactions are needed in BEC, but understanding BEC on the micro level still remains a problem [1b]. Here, the UEOS, if found, may come to the rescu [1a]. A universal equation of state must also give a universal molecular potential energy curve (PEC),

---

[1] For a synopsis of pre-quantal theoretical physics and the classical basis of the Van der waals equation, see [2b].

whereby information on bonds derives from the interaction between atoms. The number of parameters for a PEC must be small to warrant the universal character of the UEOS [3-4,6-7]. Despite its simplicity and against expectations, a simple common sense -1/r law is the most likely and natural candidate for generating PECs in a universal way [3-4]. A -1/r law must therefore be an essential part of the UEOS, just like it is in BEC with the mean field approximation and the *Gross-Pitaevskii* (GP) equation [1b,5]. However, on the micro level, the price for this simplification in chemistry seems high if the -1/r law is of Coulomb type: *intra-atomic charge-inversion*, with natural anti-atoms and atom *handedness* appear [1a,4]. This conclusion is nevertheless in line with the H-line spectrum [1a,8].

Historically, common sense energy lowering continuously varying -1/r attractions (Newton, Coulomb) have always been predominant in bound state theories for natural stable systems, despite the appearance of discrete quantum effects at the micro level. Nevertheless, it is too simple to say that quantum effects break down the classical mechanics of the macro-world. The reason is that *the most important phase transition of all in nature between left and right, which is invariantly observed in both micro- and macro-worlds, remains an open problem even for sophisticated quantum (field) theory.* As a consequence, it is likely to suppose that a not yet completely understood left-right transition mechanism must be incorporated in some way into the real UEOS. In this work, we give a semi-quantitative and phenomenological proof for this possibility.

In fact, the failure of modern theory (QFT) to deal conclusively with left-right differences in micro-system H is proved by the mere fact that one is forced to find out *experimentally* [9,10] if the symmetry of the 1/r electron-proton Coulomb bond is broken in a transition from an *atom phase* H to a *charge inverted antiatom phase* $\underline{H}$ (*antihydrogen*). However, molecular band [1a,4] as well as atomic line spectra [8] reveal conclusively that this transition really occurs in nature, a thesis, suggested in 1985 [11]. If so, the chances are great that a left-right mechanism is also crucial for the UEOS, which should be valid for all details of all phase transitions in all natural systems, whether of micro- or macro-scale.

Classical chiral behavior (left-right transitions), known from the 19$^{th}$ century, is important for the morphology of macro-structures $M_a$, even for living species. The asymmetry in a microscopic unit $M_i$, is responsible and decisive for the shape of its macrostructure $M_a$ [12,13].

Chiral symmetry is so elementary in mathematics, where it corresponds with *the difference between left- and right-handed Cartesian 3D reference frames*, that its physical/chemical implications for the smallest and simplest stable systems in nature tend to be overlooked, if not considered as trivial too. Yet the mystery surrounding antiatom $\underline{H}$ proves that quantifying this almost trivial mathematical difference in terms of physical, natural systems like neutral elementary particles is very difficult [12]. The greatest danger thereby is that solutions be promoted by conventions, which is not the most reliable way to deal with natural processes. In fact, we recently showed that, in reality, nature does not adhere to the very stringent human conventions made on charge-distributions in electrically neutral systems [1a].

*In the context of the history of left-right differences, it is essential that the UEOS provides curves for micro and macro with the very same analyticity at critical points* [1b]. To prove this in a phenomenological way, *we start from scratch, just like we did in* [4]. *We skip the details of a full thermodynamic treatment for macro- and of a full quantum treatment for micro-phase transitions, most of which is textbook material.*

## 2. Multiplicative scaling for macro-systems

Starting from scratch for interactions in natural many-body-systems means that the first problem is the definition of a homogeneous macro-system $M_a$ as a function of a number of identical micro-systems $M_i$. The UEOS requires that (at least a large part of the) *intra-* and *inter*-particle interactions are of type $1/r$, since they are only differentiated by *arbitrary* external system-independent scale factor N, defining the *scale* of macro-system $M_a=NM_i$. In essence, N is arbitrary: it is the number of units $M_i$ we need *to observe changes in a macro-system* $M_a$ and it cannot be decisive for *physics* and/or *symmetries* within a single unit $M_i$. If N seems critical *in practice*, this is due *to the sensitivity of equipment and/or to the choice of the experiment, not to the unit. Pressure chambers or* classical *traps* impose perturbations on a system by changing *external* conditions. Natural systems, like atoms and bonds, are stable and retain their properties without a trap. Whatever the conditions imposed on trapped systems, a simple way to define macro-system $M_a$ is in a multiplicative way

$$M_a = NM_i \qquad (1a)$$

the common sense basis for Avogadro's number [2b] and metrology. Ultimately, this produces *average* properties of unit $M_i$ after scaling $M_a$ with N, since

$$M_a/N = M_i \qquad (1b)$$

It is impossible to refute the logic of (1). Historically and for centuries [2b], experiments on $M_a$ intrude in the *average* properties of $M_i$, *for which transitions between states of aggregation cannot apply.* There could be a difference between *averaged* $M_i$ and *isolated unit* U, when perturbed in the same way, but the *intrinsic* properties of a species cannot change by confining N units to a trap. Few traps serve *scientific purposes*, as is evident from many traps (reaction vessels) in industrial or applied (bio-) chemistry. This illustrates why N is not *physically important when it comes to assess critical points contained within the unit structure itself. It is a very useful practical and economic scale factor*, leading to Avogadro's number N [2b], the unique *metrological scale factor between macro and micro.* With BEC-transitions, *there must at least be one critical point in a single unit* (like a triple point), where properties are singular. Small deviations from this point can cause drastic changes in the unit's properties [4,8].

**A paradox. Stable bound 1/r BECs without trap. Antihydrogen**

This apparent paradox[2] on the *physical importance* of N and the multiplicative scaling process underlying (1) can be solved by considering a generic phase transition *within a unit* U, which must contain exactly that unit- or species-specific information needed for macro-behavior of $M_a$, in particular for a transition between states of aggregation. This is about the *self-organization[3] of stable electrically neutral units in nature, governed by the characteristics of the species itself* [13].

If both intra- and inter-particle interactions are of 1/r type, unit U must be divided in at least two complementary parts, in a real or virtual way. In terms of the EPR-paradox [14], *non-local additive* contributions in micro-system $M_i$ (by a division in internal parts) can generate an asymmetry in *naked* unit U

$$U_{A,B} = u_A + u_B = u_R + u_L \qquad (2)$$

which may lead to a *reversible and continuous* transition from a local state (phase A) to a non-local state (phase B). For *complementary* systems (2), this gives p=2 for a single component and f=1, which makes this micro-system monovariant like a liquid-gas transition for $H_2O$. Two-state description (2) also applies to BEC, a phase transition, which, *just like any other*, must be covered by the UEOS. If (2) leads to a universal phase transition in a unit atom H, the *unique* UEOS may be identified. EPR-related problem (2) is fundamental since even the smallest difference between phase A and B (like left-right symmetry differences[4] as well as fermion-boson, odd-even…), is *multiplied* by N in macro-system $M_a$. As remarked above, symmetry differences are particular: mathematically, they are simple but physically they are not [16]. Energy differences due to parity violation are cumbersome, almost unsolved exercises in atoms and molecules, although the difference between left- and right-handed reference frames is evident, if not trivial.

Even H-symmetry seems simple but it is not. H is *a composite attractive self-bound Coulomb 1/r system*: it is perfectly *symmetrical* in terms of *charges* but very *asymmetrical* in terms of *masses* [16]. For system H with mass $+m_H$, (2) can be written as a composite *complementary* mass $+m_H = m_e + m_p$, *which has to be conserved in any perturbation* [16] by changing environmental conditions. Just like with the liquid-gas transition in $M_a = N \cdot H_2O$, unit mass and molecular formula $H_2O$ are conserved during a macro phase transition. This is determined by a $1/r$ law, where r is the *average* distance between a pair of $H_2O$ molecules (the basis of MFA, mean field approximation, applying for BECs [1b,5]). In BECs, the de Broglie length is a measure for critical $r_{crit}$, where the transition occurs [5]. Since this length is related to the Bohr length, BEC phenomena are critical for the micro-macro gap too. It is therefore likely that, *even without a trap*, stable self-bound BECs exist, due to the generic nature of a $1/r$-law, be it of electromagnetic or even a gravitational nature [17].

---

[2] This N-paradox is similar to that of Einstein, Podolsky and Rose [14] on the completeness of wave mechanics.
[3] Fig. 1 in [13a] is a nice illustration of self-organization in natural structures (*supramolecular chemistry*): helical macro-structures are generated because of an asymmetry within the building block, the micro-unit.
[4] The problem is to find out whether a (multiplicative) single wave function for a system is sufficient to describe this system completely, see for instance equation (1) in EPR-paper [14]. In general, (2) leads to a wave function $\psi_{A,B} = \psi_A \pm \psi_B$ instead of a single wave function $\psi$ without subscripts to differentiate between states. Following Hund [15], intriguing states are A=L (left) and B=R (right), leading to chiral symmetry (breaking).

**Similarity of micro and macro phase transitions and left-right character of the UEOS**

Van der waals interactions $1/r^n$ are important for BEC [1b,5]. Starting from scratch here involves the ideal gas law for *a macro-system without phase transition*

$$PV = RT = NkT \qquad (3a)$$

and brings in the micro-world with Avogadro's N. With $v=V/N$, the micro-energy for a unit

$$Pv = kT \qquad (3b)$$

is of generic type U, r just like Coulomb's or Newton's $1/r$, if the force obeys $1/r^2$ [1,18]. In reduced form, this leads to a relation for a micro-system deriving from macro law (3a) or

$$1/r \sim kT \qquad (3c)$$

which must be an equation[5] for *ideal behavior in micro-systems*, given its origin (3a).

With (3a), $P=RT/V$, a macro P, V-diagram is of hyperbolic Coulomb type but, therefore, it *cannot be extrapolated to negative pressures (metastable states) observed in non-ideal or real gases, whereby condensation gives birth to another state of aggregation (liquid)*. Inverse P, 1/V-diagrams of linear Hooke type are easily extrapolated *to negative pressures*, whatever their physical meaning (see below).

For a gas-liquid transition of a real originally dilute gas (as in BECs), real macro diagrams obey the classical Van der waals' equation [2a]

$$(P+a/V^2)(V-b) = RT = NkT \qquad (4a)$$

*historically the oldest form to quantify 2-vertex models like in* (2). The two correction terms in (4a), *a* for P and *b* for V, appear by giving up the *point model*[6] for an ideal gas and replacing it with linear model (2). Like with (3a), P, V-diagrams as well as P, 1/V-diagrams can be generated with (4) with similar consequences as above for *negative pressures (and metastable states)*. A *reduced numerical* Van der waals equation is a UEOS-candidate. In the quest for the UEOS, going on for over a century, many EsOS were proposed but we cannot review these here (see [1-7]).

Rewriting (4a) in PV-form and scaling by N leads to

$$Pv(1+a/V^2)(1-b/V) = kT \qquad (4b)$$

---

[5] Since $1/r \sim 1/\lambda$, (3c) leads to ratio $h\nu/kT=(hc/\lambda)/kT$, the basis of the Planck-Einstein theory for specific heats. As such, (3c) is also the basis of all modern quantum theories and their interpretation of phase transitions.
[6] BEC observations [19] are consistent with atoms interacting at short distance like hard *spheres*.

which, with (3c), will lead to similar *analytical* deviations, including equivalent phase transitions, for micro-systems, governed by a 1/r law. These important cases are discussed below.

For *small* b, an equivalent form [2b] for macro behavior (4) can be expanded as

$$P = (RT/V)(1 + b/V + \ldots) - a/V^2 \qquad (5a)$$

where the linear term in 1/V, *representing ideal behavior* (3), is adjusted with two quadratic terms in $1/V^2$ of opposite sign, *of which only one, $RTb/V^2$, is temperature dependent*. With (5a), deviations $\Delta$ from ideal behavior *for small b* are given by

$$\Delta = P - RT/V = bRT/V^2 - a/V^2 \qquad (5b)$$

*Only if both a and b are very small, $\Delta$ tends to zero and both the macro- (3a) as well as the derived micro-system (3b) will behave almost ideally.* Although a transition between states of aggregation is easily visible, one expects that left-right transition within a single unit (translated in energy Pv) is not.

This is illustrated with the P,V-diagram of Van der waals type (4), given in Fig. 1a. The most characteristic but also most difficult point is the intermediate maximum for P at a critical particle separation governing the phase transition (and the so-called metastable states and negative pressures). The analyticity requirements for the (U)EOS at this turning point are extremely important [1b], since this point announces a switch from *positive to negative pressures*.

Without loosing accuracy or information, P, 1/V-diagrams can be constructed with the same data as shown in Fig. 1b. Although the mathematics is simple, the difference between V- and 1/V-diagrams is large[7] and this must have physical implications for observed macro phase transitions. The complex Van der waals-curve for a liquid-gas transition (Fig. 1a) with *its critical P-maximum during the phase transition* and *its negative pressure domain*, transforms into a Mawellian *binodal* (Fig. 1b). *This analysis shows that phase transitions in the macro-world, obeying a Van der waals-Maxwell binodal, are similar to phase transitions observed in the micro-world, obeying a Mexican hat or double well curve of Hund-type* [15]. These apply to enantiomers, deriving from 19[th] century work on *chiral structures* [12]. Hund's work [15] is consistent with tunneling effects, typical for double well potentials. *All these features are important for BEC observations too* [5,19].

*At this stage, the important conclusion from Fig. 1a and 1b is that Hund-type micro phase transitions correspond with the generic difference between stable left- to right-handed forms (enantiomers) of neutral unit U or $M_r$. But these are of the very same type as the macro transitions, obeying pre-quantal Van der waals' equation (4). Going from a V- to a 1/V-representation for phase transitions in macro-systems reveals a generic mechanism, typical for left-right (a)symmetry in stable micro-systems, as suggested by (2).*

---

[7] Fitting data points in P, V-curves is difficult. For P, 1/V-curves fitting is rather smooth [18].

**Chiral (a)symmetry and generic natural symmetry breaking in the UEOS**

*The similarity of macro- and micro-phase transitions revealed by Fig. 1b and Hund-type Mexican hat curves* [15] *means that chiral (a)symmetry is probably the unique basis for the UEOS, if it exists* [1a]. *All experimental evidence used thus far is pre-quantal, as it derives from seminal work of the 19$^{th}$ century (i) on transition between classical states of aggregation and (ii) on optical activity of stable neutral but asymmetrical chiral structures.*
Chiral symmetry in 4D space time transforms (-x,+y,+z,+t) in (+x,+y,+z,+t). For numbers, parity is an exact unbroken symmetry, an algebraic 1D-effect on a field-axis. If so, a critical point, characterizing a phase transition understood with a 1/r interaction, must be placed between separations smaller or greater than critical separation $r_{crit}$. For this to be applicable to a *single* unit, a similar interaction must apply to for its two complementary parts A and B or R and L (2). To arrive at parts, a unit must be divided[8] as in (2). In either case, *micro or macro*, interactions obey $r_A < r_{crit} < r_B$, where suffixes A and B stand for two units 1 and 2 in the macro-world, for 2 phases, for 2 states or for complementary parts A and B in the micro-world, corresponding with say local and non-local, left and right… states as in (2). When 2 $H_2O$ molecules in the gaseous phase, forming a *unit* of size $(H_2O)_2$, are at a distance $r_{crit}$, a liquid phase can set in. If a 1/r-law is required, sequence $1/r_{A(B)} > 1/r_{crit} > 1/r_{B(A)}$ is typical for a universal phase transition and, hence, for the analyticity of the UEOS [16]. Inverse diagrams (Fig. 1b) lead to a quantitative link between the observed macro-transition between states of aggregation in $M_a = NM_i$ and a similar, internal micro left-right transition between states (2) in unit system U or $M_i$, *pending a perturbation*. Only a transition within unit U can explain why macro-transitions between states of aggregation are *species-specific*. To make sense, unit $H_2O$ must have a critical internal separation $r_{crit}$, which governs the liquid-gas transition. If all systems exhibit this behavior, *a numerical physics- or species-independent UEOS must exist and be valid for micro- and macro-systems. Then, it will have to be applicable for BEC as well as for any phase transition in nature.* This apparent connection with BEC is sufficient to arrive at a quantitative, analytical assessment of the direct link between macro- and micro-behavior.

**Quantitative micro-macro link using BEC and the de Broglie relation**
With (3b), Pv must be of type 1/r in the micro world, as in (3c). However, *with radiative instead of thermal excitation*, kT must be replaced with $h\nu = hc/\lambda$, leading to the equivalent of (4b) of form $1/r \sim hc/\lambda +$ *correction terms*, of exactly the same form as those prevailing for macro-systems and made explicit in (4b).

The inverse form of (3c) for micro-systems is of type $r \sim \lambda/hc$ like in the de Broglie equation, which places a constraint of the separation between 2 units (macro) or 2 parts (micro), as in (2) and the wavelength of the incident perturbing electromagnetic radiation. This is in line with the fact that, in BEC, the de Broglie-*length* is a measure for $r_{crit}$ [5]. A de Broglie-*length* is linked with a Bohr-*length* (atom radius) by means of *the standing wave equation*

$$2\pi r = n\lambda \qquad (5c)$$

---

[8] Of course, this places constraints on the definition of a natural stable *unit*, a problem we cannot deal with here.

*the basis of Schrödinger's wave mechanics.* This is exactly as expected on the basis of (3c) but an explicit scale factor is needed before one can proceed. Since symmetries reflect field effects, comparing *lengths* (separations) is to be recommended over comparing a circumference and a (wave) length as in (5c). An *exact* alternative for (5c) is therefore

$$2r/\lambda = n/\pi \qquad (5d)$$

wherein *a length, the diameter (not the circumference) of a Bohr system* is confronted with electromagnetic radiation $h\nu$, typified by its wave *length* $\lambda$, the sole basis for resonance based on field effects. With (3c), we expect $r/\lambda \sim 1/hc$ for the ideal behavior of a micro-system.

In (5d), a quantitative connection between principal quantum number n and irrational number $\pi$ appears, which needs to be validated [20]. If correct, an angle of 180° degrees can be interpreted as the end result of a permutation, like that of 2 unit charges in a Coulomb interaction: $e_1e_2$ and inverse $e_2e_1$ with a mirror plane placed in between them at exactly ½$\pi$ [8,18]. This leads to more generic information on states A and B or R and L in (2): *after a permutation of 2 unit charges by exactly 180° or $\pi$ radians*, Bohr atom H ($e^-$, $p^+$), say state R, transforms in its *mirrored, charge-inverted, anti-atomic (antihydrogenic)* equivalent $\underline{H}$ ($e^+$,$p^-$), state L [8,16].

The field scale factor, missing in the de Broglie equation (5c), is needed explicitly to interpret (3c) and (4b) on the basis of spectral information on unit U. We rewrite electromagnetic energy as

$$h\nu = hc/\lambda = 2\pi\hbar c/\lambda = (2\pi/\alpha)(e^2/\lambda) \qquad (5e)$$

where $\alpha$ is the fine structure constant, a dimensionless scale factor $e^2/\hbar c=1/137,03566675$, first introduced by Sommerfeld. An additional *multiplicative numerical scale factor* S, like N in (3) but overlooked therein, is needed to convert electromagnetic energy $h\nu$ into *its closest Coulomb equivalent* $e^2/\lambda$. This field scale factor S is competitive with scale factor N above. S is needed when we change from perturbation by heat (3b) to perturbation by radiation. As easily verified, (5e) is numerically given by $2\pi/\alpha$, or

$$S = 2\pi/\alpha \approx 861 \qquad (5f)$$

Strangely enough, this is not too far [20] from (inverse) classical recoil

$$S' = \tfrac{1}{2}m_p/m_e \approx 918 \qquad (5g)$$

*Numerical multiplicative scale factors S and S' are neglected in the de Broglie relation between wavelength and Bohr-length* (5c). We obtain *a field-corrected de Broglie relation for* (5d), whereby either the fine structure constant (5f) or recoil-like (5g) acts as a scale factor

$$2r/\lambda = 2(r/e^2)(e^2/\lambda) \sim (n/\pi)/S \qquad (5h)$$

expressing the well-known fact that electromagnetic energy has a different asymptote than Coulomb's [20]. The ratio is about 1000, see (5f) and (5g) but it must be acknowledged that scale factor S (or S') is competitive with N when we go from micro to macro. The asymptotes can be linked, through the de Broglie equation, with a system's *particle properties* (Coulomb field) or its *wave properties* (electromagnetic field). In quantum theories like Bohr's, the asymptote difference is incorporated in *a multiplicative way* in the electron's angular velocity ($\alpha c$). An *additive* mass correction reduces all energies slightly by $1/(1+m_e/m_p)$, showing that both (5f) and (5g) are important for metrology (the basis for a micro-macro distinction). The effects above obviously relate to radial rather than angular velocities, whence they are field effects as expected for left-right differences. *If so, chiral field effects are invisible in standard or original Bohr H-theory, wherein only angular velocities are quantized with principal quantum number n (Bohr's famous quantum hypothesis).* Asymptote differences, or the competition between numerical scale factors S and N, can make the detection of a phase transition in simple unit H difficult if observed spectra are used. If ideal behavior is linked with one asymptote, non-ideal behavior can be linked with the other one, reflecting, finally, the effect of the magnitude of the terms in a and b in the classical Vanderwaals-Maxwell analysis (4)-(5) on the deviations from ideal behavior. If both a and b in (4b) are very small for a unit like H, detecting a phase transition within H may be difficult indeed, since also the interference of scale factor S or S' over N must be accounted for.

*The only generic aid is that symmetry implications are evident and unambiguous.* When comparing Coulomb and electromagnetic H-fields with (5e), it is evident that the latter is more symmetrical (*achiral*) with respect to 2 unit charges than the less symmetrical (*chiral*) first for H, where 2 unit charges reside on 2 particles with very different mass (electron and proton). *Whereas we expect electromagnetic energy (radiation) $h\nu=hc/\lambda$ will not be sensitive to a left-right difference, a Coulomb energy for H, determined by $e_1e_2/r$ can be left-right sensitive because of the large mass difference of its two complementary parts.* Since these 2 energies are linked to different asymptotes because of S or S', one should be cautious in the choice of the asymptote to reveal chiral behavior of system H, if any. To do so, one should focus on *chiral* rather than *achiral* energies, since the latter are, by definition, insensitive to a left-right difference. This distinction can now be applied to system H on the basis of earlier results [4,8,21]. *We remind that with (1), the Van der waals-equation (5a) for macro phase transitions is simply reduced to its micro equivalent by replacing V by v and R by k, which cannot alter the physics involved. Only the scale is reduced: macro must behave exactly like micro, the result of comparing the curve in Fig. 1b with Hund-type curves.*

**Achiral formula for simple natural system H and errors due to chiral behavior**

Chiral effects in the Coulomb H-bond may be due to *an internal phase transition* possible with (2), when H is perturbed. Its contributions must be due to chances in *radial velocities*, left uncovered in standard Bohr theory, as argued in the preceeding section. A Mexican hat curve detected in the H-line spectrum confirms this thesis in more detail [21]. Surprisingly or not, this curve has a critical point n=π, exactly as derived in (5d) and, therefore, seems to meet all of the above generic criteria for phase transitions observed in natural systems (small or large, micro or macro, classical or quantum) and the UEOS behind them.

Since it cannot account for left-right differences analytically [16,18,20], Bohr's theory can only be classified as *achiral*. Then, Bohr's *achiral* $1/n^2$ theory[9] becomes a substitute for ideal (rotator) behavior of the electron-proton bond, since it does not allow phase transitions, just like the ideal gas law (3a-b). Both approaches predict continuous system behavior for micro systems like H (Bohr theory) and macro systems, like ideal gases. Neither allows for any phase transition.

It is known for long that simple Bohr theory, even for H-ns-states, is not exact, although its errors are relatively small (order 0,01 cm$^{-1}$ or 10$^{-6}$ eV). But, exactly due to its simplicity, it is very simple to calculate the errors of achiral Bohr theory, when compared with observed data [8]. This is a straightforward micro-application of macro-equation (5b).

But since Bohr theory is *achiral*, only its errors can contain *signatures for left-right differences in system H* [16]. And since a left-right transition in H is connected with parity violation, *it can be considered as the most generic phase transition of left-right type occurring in nature*. Now, the existence of a left-right term in the UEOS depends on the type of the errors produced by achiral Bohr theory, probably deriving from neglected radial field effects (see above).

Proving that the achiral nature of Bohr theory for hydrogen is elemental, since, if H is charge-inverted, it becomes antihydrogen H̲. In both cases, Bohr theory gives identical Hamiltonians ***H***

$$\boldsymbol{H}(H) \equiv \boldsymbol{H}(\underline{H}) \equiv \tfrac{1}{2}\mu v^2 - e^2/r$$

and *identical eigenvalues* (energies for level n) using Rydberg R

$$E_n(H) \equiv E_n(\underline{H}) \equiv -R/n^2 \qquad (5i)$$

*which proves Bohr's theory is achiral by definition*. And if so, chiral effects can only show through radial effects, unjustly neglected by Bohr.

If the errors of Bohr H-theory would be at random, no signature for a left-right difference in H can be detected. *But, as soon as these errors are of binodal or Mexican hat type like Fig. 1b, a left-right difference in Coulomb electron-proton system H, as well as the corresponding phase transition, is proved by the H-line spectrum.*

---

[9] The qualification *achiral* means here that Bohr theory cannot differentiate between left and right or between a Coulomb interaction $e_1e_2$ and its inverse $e_2e_1$ after a permutation/inversion of charges in neutral unit H [16].

Moreover, *an expansion (excitation or perturbation) of system H in function of principal quantum number n (from n=1 to n=∞) is similar with the classical expansion of a macro-system $M_a$ or N units $M_i$ in function of the available volume V in a pressure chamber.*

For the Lyman ns-series of system H, Erickson QED-data [22], accurate to $10^{-7}$ cm$^{-1}$ and Kelly data [23], accuracy $10^{-4}$ cm$^{-1}$, are available. Since Erickson-data were used earlier [8,21], we now use Kelly's to show that high precision is not needed to detect a left-right transition in natural H. Errors with *achiral* Bohr theory, *now promoted to the status of chiral effects in natural system H* [16], refer its ground state n=1 with Kelly's H-limit $-E_{1H} = -R_H$ = 109678.7737 cm$^{-1}$. With (5i), the errors are

$$\Delta E_{nH}(R_H) = E_{nH} + 109678.7737/n^2 \quad \text{cm}^{-1} \quad (6a)$$

With the harmonic Rydberg [8], equal to $R_{harm}$ = 109679.3589 cm$^{-1}$ *with Kelly-data*, errors due to chiral behavior of H are

$$\Delta E_{nH}(R_{harm}) = E_{nH} + 109679.3589/n^2 \quad \text{cm}^{-1} \quad (6b)$$

The difference between (6a) and (6b) is a very simple, linear and rather small asymptote or Rydberg shift of only 0,58 cm$^{-1}$ applied to Bohr formula (5i) to describe the same H-system. One could even expect that this very small, almost insignificant asymptote shift ($5.10^{-6}$) would not even be important in terms of physics. This *expectation* is however, not met.

**Results: Signatures for a left-right phase transition in the line spectrum of unit H**

The results for micro system H, obtained with (6a) and (6b), are presented in Fig. 2a and 2c, where the errors of Bohr theory are plotted versus n. Fig. 2b an 2d give the same errors plotted versus 1/n, to retain the similarity of the treatment of the macro phase transition in Fig. 1a-b. It is evident that, throughout the complete Lyman-series, the errors of *achiral* Bohr theory are not at random at all [1a,8,16,21]. On the contrary, they follow exactly the same pattern as observed for a classical macroscopic phase transition like those described by Van der waals-Maxwell type curves, given in Fig. 1a-b.

*The almost analytical correlation between the two sets of results for macro-behavior in Fig. 1a-b and for micro-behavior of simple unit H in all Figs. 2 is too striking to be coincidental.* The only conclusion is that natural system H can exhibit an internal phase-transition probably due to charge-inversion, not covered by Bohr theory (5i). *Hence, the 2 natural phases allowed for neutral species hydrogen are hydrogenic H and antihydrogenic H, as argued above. By virtue of this striking similarity, we could already conclude at this stage that the observed phase transitions in micro systems obey the same laws as those known since the 19$^{th}$ century for macroscopic phase transitions.*

However, due to the spectral details in the micro-case (impossible to arrive at in the macro-world), it is possible to unravel the mechanism behind the micro phase transition in H. Fig. 2a-d show the influence of the small asymptote shift (*asymptotic freedom* [1a,3]) on the shape of the curve describing the phase transition and the relative depths of the two wells. Looking at Fig. 2d, only the harmonic asymptote $R_{harm}$ in (6b) provides a true Mexican hat curve for *the two H-states (2) allowed in nature but hidden in its spectrum. Comparing Fig. 2b with 2d shows that the asymptote defines the shape of the curve. In Fig. 2b for instance, the Bohr Rydberg in (6a) focuses the attention upon only one well, suggesting that a one state Bohr model for H is indeed a good approximation. With Fig. 2b, the 2 wells are apparent, showing that Bohr's one well model, implicit in (5i) does not suffice.* This is a confirmation of the EPR-paradox and a validation of model (2) instead of the classical point model.

After comparing the results illustrated in Fig. 1-2, there is much more consistency than hitherto believed or even imaginable between (a) 19$^{th}$ century Vanderwaals-Maxwell theories with spinodals/binodals for transitions between states of aggregation in macro-systems $M_a=NM_i$ and (b) internal phase transitions as observed in simple unit H, *the so-called simple electron-proton Coulomb bond*. This places question marks on the validation of highly sophisticated bound state QED [24] for micro-systems $M_i$ like H [8,18,21]. Confronting micro with macro leads to novel results for *natural generic chiral symmetry breaking*, involving enantiomers H and H̲, not even in reach by bound state QED [24] despite its sophistication, as argued before [8,16].

**A quantitative explanation: density fluctuations and critical densities**

The observed phase transition in system H, illustrated in Fig. 2, must find its origin on the field axis (see above). Hence, a plausible explanation for these modifications of the system's size in function of principal quantum number n is in terms of density fluctuations of the total system, while the absolute total mass $m_H$ of system H ($m_H$ = 1837,1526675 $m_e$) is conserved throughout. Using the de Broglie relation, critical densities can occur for principal quantum number n scaled by π, see (5d). Quantitative implications of density fluctuations are discussed elsewhere [27]. We also found that also higher Z one electron systems exhibit the same behavior [27].

**Quantitative implications for the UEOS and for symmetry breaking in natural systems**

It appears that getting *at signatures for generic left-right symmetry breaking in natural neutral systems like H does not prove to be really difficult*, despite common expectations, which led to the experiments in [9,10]. Rewriting (4b) for micro system H, perturbed in a radiative field

$$1/r \sim S(e^2/\lambda)(1 + b/V+...) - a/V^2 \qquad (7)$$

is obeyed fairly enough by available H-data (see Fig. 2). Even the details of micro system H are almost congruent with the details of a real or natural macro system, as described by the Van der waals-Maxwell approach (see Fig. 1), a surprising result after all. This result is, evidently, due to the fact that constant a in (7) is related to *the dielectric constant of a species*, as remarked early by Maxwell when he heard of the Van der waals-equation.

But we also arrived at three additional novel results, important for a number of reasons, including the UEOS.

(i) For its binodal or Mexican hat curve to show (Fig. 2b and 2d), observed H-data must be extrapolated to an *inexistent outer left to uncover the complete double well curve for chiral H*. But at least, this is now easily understood classically, just like Van der waals' curves in Fig. 1a-b. At the minimum of the left well in Fig. 2d, system H simply disappears: here, atom H transforms in 2 sub-systems, an electron and a proton at infinite separation, which, from this point on, start a life of their own. The separation or H-volume is so large, one can no longer speak of *unit system* H. Electron and proton remain nevertheless intimately connected (*entangled* in EPR-terms) as they remember very well their common origin, unit H. This is obvious by the inverse process: bringing any electron together with any proton from infinite separation will automatically and invariantly form unit system H.

The macro equivalent in Fig. 1b is that here the available volume V has become much too large for a single unit to be measurable: here, the species has *disappeared* virtually from the natural measurable scene by *dilution*[10] (*infinite average separation*). In order to view the complete symmetry picture, to which a macro system obeys in terms of measurability, only extrapolation to the not measurable *inexistent* left reveals the complete symmetry curve (*macro Vanderwaals-Maxwell binodal*). An unavoidable extrapolation to the left acts like a *trompe-l'oeil*, which hides the underlying chiral symmetry [16] for both micro and macro and their equivalence if not identity. Mathematically, this corresponds with the fact that the ideal Coulomb law, as an inverse power law, is confined to a single world and cannot even be extrapolated to another one (where another well takes over). But going over to more natural variable ($\pi/n-1$) as in [21] easily removes most of the objections towards extrapolation procedures of this kind.

(ii) A second novelty for symmetry breaking in micro-systems is a very specific question about the UEOS we must still identify. This question is overlooked in bound state QED, except for Bethe logarithms [24], needed to account for Lamb shifts.

From a comparison of Fig. 2a for micro H and Fig. 1a for macro, it is obvious that there is a critical $n_{crit}$ in natural system H at $5< n_{crit} <6$. This is remarkable since it is *quantitatively linked with the Bohr H ground state* $R_{1H}$ of 109678.7737 cm$^{-1}$, abandoned in bound state QED as well as in current metrology in favor of limit $R_{\infty H}$. The latter is the basis for the absolute Rydberg, defined as $R_{\infty H}(1+m_e/m_p)$, the most important constant for metrology of the (micro) quantum world [25,26]. *Bohr's H ground state $R_{1H}$ is needed explicitly and justly to arrive at a curve for H (Fig. 2a), similar to a conventional Vanderwaals-Maxwell P,V-diagram for macro phase transitions (Fig. 1a).* This maximum at $n_{crit}$ in Fig. 2a for expanding /compressing H plays the same role as the relative P-maximum in the P, V-diagram in Fig. 1a for boiling water or condensing steam. In either case, these intermediary critical points cannot but play *a fundamental role in natural generic symmetry breaking processes between two allowed stable phases of the same species*, connected with perturbing natural neutral micro-systems like H and $H_2O$ as in (2) and macro-systems like NH and $NH_2O$ due to multiplicative scaling (1). In the end, our work justifies the EPR-thesis be it on experimental grounds not even thinkable at the time their thesis was published.

---

[10] In [18], this was described as '*et le combat cessa faute de combattants*'.

This explicit quantitative focus on another *physical* critical n-value in system H, different from the critical n-values applying for *mathematical* left-right symmetries or permutations/inversions [16], equal to ½π [8] and π [8,21] respectively, cannot but lead to *a generic physical symmetry breaking operation, completely absent in any theory for H thus far.* Quantifying this will bring us directly to the UEOS. *We therefore expect that only the unique competition between mathematical and physical symmetry breaking mechanisms, operating in system H, will reveal why n=1 corresponds with the ground state of H.* Although finding from first principles the origin of this equally important novel critical *physical* n-value between 5 and 6 for system H seems difficult (if not superfluous after having exposed already the *mathematical* critical n-values for left-right symmetry in the same system H [8,21]), this exercise must be done in the advent of further BEC- and *artificial antihydrogen* experiments [27].

(iii) A $3^d$ novelty is rather spectacular, in that the micro phase transition is available with much more experimental detail than in the macro-case. As shown by Maxwell, the surface of the curve above the constant pressure line (during the process of boiling or condensing) must be exactly equal to that of the curve below the constant pressure line. This constraint follows from the fact that it is *simply impossible to measure negative pressures*, although implicit and predicted by the Van der waals-curve (see Fig. 1a). This uncertainty about the negative pressure domain disappears for the micro phase transition in species hydrogen. In fact, 5 accurate data points are available below $n_{crit}$ between 1 and 5 (see Fig. 2a). This is quite remarkable, as it would allow, for the first time ever, studying experimentally negative pressure domains, accompanying phase transitions. *H spectral data could even be used to test the validity of the Maxwell prediction for a phase transition.*

Concrete and imminent consequences of these surprising details of the H-line spectrum, illustrated in Fig. 2a-d, for the ongoing CERN-AD experiments on antihydrogen [9,10] are in [27].

**Conclusion**

The striking similarity between critical behavior in macro and micro as exposed here in great detail, was left unnoticed for almost a century, although it is at the basis of Avogadro's hypothesis (metrology). *The* UEOS must exist. A few analytical benchmarks (critical points) for the UEOS are that it must be of left-right or chiral type, since it is the unique, generic and system-independent solution to deal with the physically ambiguous but mathematically unambiguous difference between left- and right-handed Cartesian reference frames. As shown in [21], a simple cosine law automatically gives benchmarks ½π (for a mirror plane) and π (for a permutation or charge inversion), *two elementary signatures for chirality, easily detected in prototype system H as we showed here* [8,16,20,21].

Natural H, the so-called simple electron-proton Coulomb bond, is not a single well but a double well system. Current micro *artificial antihydrogen experiments* [9,10] should be placed in this broader perspective, with due *focus on and respect for* historically important, pre-quantal 19$^{th}$ century macro results on chiral behavior and phase transitions [2], as illustrated in [27]. Metrological errors deriving from of a single well H-model will have to be corrected (the absolute Rydberg may have to be reviewed). *The analytical link between micro and macro phase transitions detected here is provided with BEC and de Broglie standing wave equation.* The UEOS is at sight, since, at last, we finally know what to look for.

**References**
[1] (a) G. Van Hooydonk, Eur. Phys. J. D, OnlineFirst (2005); (b) E. Brosh, G. Makov and R.Z. Shneck, J. Phys. Condens. Matter **15**, 2991 (2003); (c) B. Gao, J. Phys. B: At. Mol. Opt. Phys. **37**, L227 (2004)
[2] (a) J. D. Vanderwaals, *Over de continuïteit van de gas- en vloeistoftoestand*, Ph. D. thesis, Leiden, 1873; (b) E. Bloch, *Théorie cinétique des gaz*, A. Collin, Paris, 1921
[3] G. Van Hooydonk, Eur. J. Inorg. Chem, 1617 (1999)
[4] G. Van Hooydonk, Spectrochim. Acta A 56, 2273 (2000); physics/0003005; physics/0001059
[5] F. Dalfovo, S. Giorgini, L.P. Pitaevskii and S. Stringari, Rev. Mod. Phys. **71**, 463 (1999); A. Leggett, Rev. Mod. Phys. **73**, 307 (2001); J.O. Andersen, Rev. Mod. Phys. **76**, 599 (2004). For chemical aspects of BEC, see W. P. Reinhardt and H. Perry in: E.J. Brändas and E.S. Kryachko (Editors), *Fundamental World of Quantum Chemistry: A Tribute to the Memory of Per-Olov Löwdin*, Vol II, Kluwer, 2003
[6] Y.P. Varshni, Rev. Mod. Phys. **29**, 664 (1957)
[7] J. Graves and R.G. Parr, Phys. Rev. A **31**, 1 (1985)
[8] G. Van Hooydonk, Phys. Rev. A **66**, 044103 (2002); physics/0501144
[9] M. Amoretti et al. (ATHENA collaboration), Nature **419**, 456 (2002)
[10] G. Gabrielse et al. (ATRAP collaboration), Phys. Rev. Lett. **89**, 213401 (2002); 233401 (2002)
[11] G. Van Hooydonk, Theochem –J. Molec. Struct. **121**, 45 (1985)
[12] M. Quack, Angew. Chem. Int. Ed. **40**, 4195 (2002); **41**, 4618 (2002)
[13] (a) V.G. Machado, P.N.W. Baxter and J.M. Lehn, CPS: inorgchem/0107001, J. Braz. Chem. Soc. **14**, 777 (2003); (b) J.-M. Lehn, Rep. Prog. Phys. **67**, 249 (2004)
[14] A. Einstein, B. Podolski and N. Rosen, Phys. Rev. **47**, 777 (1935)
[15] F. Hund, Z. Phys. **43**, 805 (1927)
[16] G. Van Hooydonk, Eur. Phys. J. D, Onlinefirst (2005); DOI:10.1440/epjd/e2005-00028-6
[17] D. O'Dell, S. Giovanazzi, G. Kurizki and V.M. Akulin, quant-ph/9912097
[18] G. Van Hooydonk, CPS: physchem/0308001
[19] M. H. Anderson, J.R. Ensher, M.R. Matthews, C.E. Wieman and E.A. Cornell, Science **269**, 198 (1995); C.C. Bradley, C.A. Sackett, J.J. Tollett and R.G. Hulet, Phys. Rev. Lett. **75**, 1687 (1995); K.B. Davis, M.-O.

Mewes, M.R. Andrews, N.J. Van Druten, D.S. Durfee, D.M. Kurn and W. Ketterle, Phys. Rev. Lett. **75**, 3969 (1995)
[20] G. Van Hooydonk, CPS: physchem/0401001
[21] G. Van Hooydonk, Acta Phys. Hung. **19**, 385 (2004); preprint in M. Koniorczyk and P. Adam (Editors): *Proceedings of the Wigner Centennial*, Pecs, 2002, http://quantum.ttk.pte.hu/~wigner/proceedings; physics/0501145
[22] G.W. Erickson, J. Phys. Chem. Ref. Data **6**, 831 (1977)
[23] R.L. Kelly, J. Phys. Chem. Ref. Data **16**, Suppl. 1 (1987)
[24] M.I. Eides, H. Grotch and V.A. Shelyuto, Phys. Rep. **342**, 63 (2001); hep-ph/0002158
[25] T. Kinoshita, Rep. Prog. Phys. **59**, 1459 (1996); B. Cagnac, M.D. Plimmer, L. Julien and F. Biraben, Rep. Prog. Phys. **57**, 853 (1994)
[26] P.J. Mohr and B.N. Taylor, Rev. Mod. Phys. **72**, 351 (2000)
[27] G. Van Hooydonk, physics/0502074; unpublished results

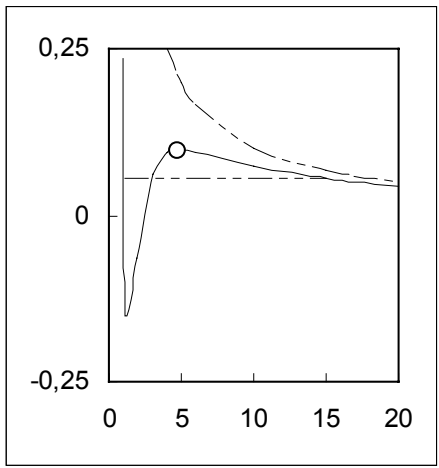

Fig. 1a P,V-diagram for liquid-gas transition (full curve: aid to the eye; dashed hyperbola: ideal gas law, dashed line: constant pressure, circle: maximum pressure at longer range)

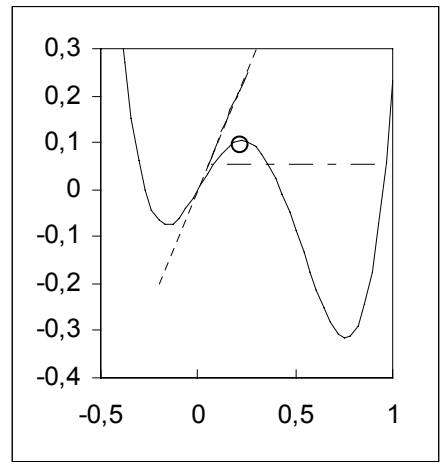

Fig. 1b P, 1/V-diagram (same data as in Fig. 1a) (full curve: 4th order fit; small dashes straight line: ideal gas law)

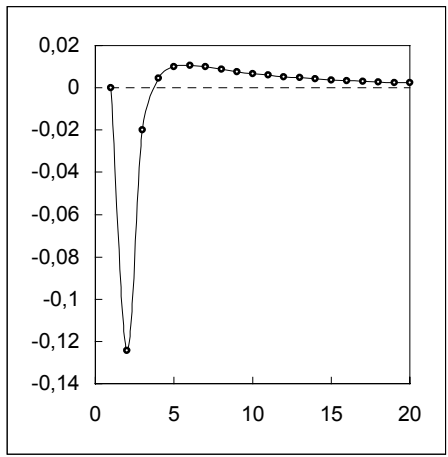

Fig. 2a Kelly data: Results of (6a) for Lyman ns-series with Bohr ground state $R_{1H}$ vs. n (curve aid to the eye)

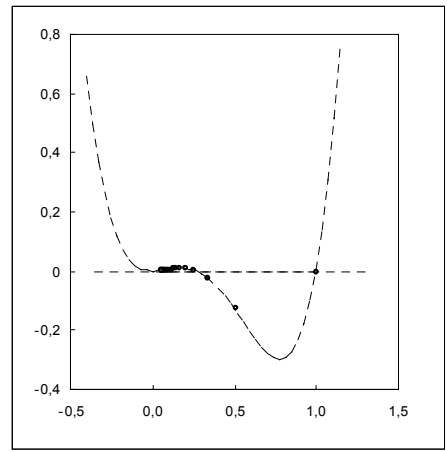

Fig. 2b Same results as in Fig. 2a vs. 1/n (curve 4th order fit, extrapolated)

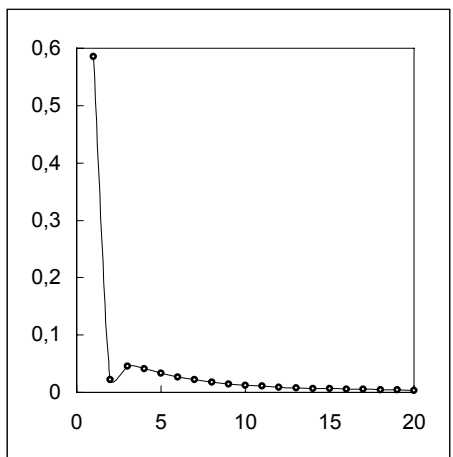

Fig. 2c Kelly data: Results of (6b) for Lyman ns-series with harmonic Rydberg $R_{harm}$ vs. n (curve aid to the eye)

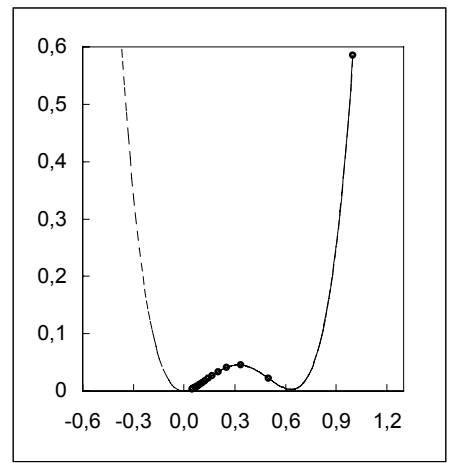

Fig. 2d Same results as in Fig. 2c vs. 1/n (curve 4th order fit, extrapolate